\documentclass[aps,twocolumn,nofootinbib,longbibliography]{revtex4-1}
\usepackage[babel]{csquotes}
\usepackage{graphicx}
\usepackage{amsmath,amssymb}
\usepackage[colorlinks,citecolor=blue,linkcolor=blue,urlcolor=blue]{hyperref}
\usepackage{mathrsfs}
\usepackage{enumerate}
\def\be{\begin{equation}}
\def\ee{\end{equation}}

\begin{document}

\title{The strange equation of quantum gravity}

\author{Carlo Rovelli}

\affiliation{Aix Marseille Universit\'e, CNRS, CPT, UMR 7332, 13288 Marseille, France\\
Universit\'e de Toulon, CNRS, CPT, UMR 7332, 83957 La Garde, France.}

\date{\small\today}

\begin{abstract}
\noindent  Disavowed by one its fathers, ill defined, never  empirically tested, 
the Wheeler-DeWitt equation has nevertheless had a powerful influence on 
fundamental physics. A well deserved one.
\end{abstract}

\maketitle
    
\section{Introduction}

\noindent One day in 1965, John Wheeler had a two hours stopover between flights at the Raleigh-Durham airport in North Carolina. He called Bryce DeWitt, then at the University of North Carolina in Chapel Hill, proposing to meet at the airport during the wait.  Bryce showed up with the Hamilton-Jacobi equation of general relativity, published  by Asher Peres shortly earlier \cite{Peres1962}:
\be
  \left(q_{ab}q_{cd}-\frac12 q_{ac}q_{bd}\right)\frac{\delta S[q]}{\delta q_{ac}}\frac{\delta S[q]}{\delta q_{bd}}-\det{q}\, R[q]=0.  \label{peres}
\ee
$q_{ab}$, with $a,b=1,2,3$ are the components of a 3d spatial metric, $R[q]$ its Ricci scalar, and $S[q]$ a Hamilton-Jacobi functional. Bryce mumbled the idea of repeating what Schr\"odinger did for the hydrogen atom: getting a wave equation by replacing the square of derivatives with ($i$ times) a second derivative---a manner for undoing the optical approximation.  This gives \cite{DeWitt:1967yk,Wheeler:1968}
\be
  \left[\left(q_{ab}q_{cd}-\frac12 q_{ac}q_{bd}\right)\, \frac{\delta }{\delta q_{ac}}\,\frac{\delta }{\delta q_{bd}}+\det{q}\, R[q]\right]\Psi[q]=0  \label{WdW}
\ee
for a functional $\Psi[q]$ to be interpreted as the wave function of the gravitational field $q$.   Wheeler got tremendously excited (he was often enthusiastic) and declared on the spot that \emph{the} equation for quantum gravity had been found. This is the birth of  eq.\,\eqref{WdW} (\cite{Dewitt-Morette2011}, pg 58).    For a long time Wheeler called it the ``Einstein-Schr\"odinger equation", while DeWitt called it the ``Wheeler equation", or "that damn equation", without hiding his unhappiness with it it.  For the rest of the world it is the Wheeler-DeWitt, or ``WdW", equation. 
 
It is a strange equation, full of nasty features. First, it is ill-defined. The functional derivatives are distributions, which cannot be squared without yielding divergences. Concrete calculations, indeed, tend to give meaningless results: strictly speaking, there is no equation. Second, the equation  breaks the manifest relativistic covariance between space and time: quite bad for the quantum theory of general relativity.  Third, there is no time variable in the equation, a puzzling feature for an equation with the ambition to be the \emph{dynamical} equation of quantum spacetime. This has raised endless confusion, and prompted a long lasting debate on the nature of time. 

Furthermore, a Schr\"odinger-like equation is not sufficient to define a quantum theory: a scalar product  is needed to compute expectation values.  A scalar product suitable for the WdW equation is far from obvious, and rivers of ink have been wasted to discuss this issue as well.  Last but obviously not least, not a single empirical observation has so far directly supported the physical correctness of the equation, half a century after its publication.  

Still, in spite of all these limitations, the WdW equation \emph{is} a milestone in the development of general relativity.  It has inspired a good part of the research in quantum gravity for decades, and has opened  new perspectives for fundamental physics.   Much of the original conceptual confusion raised by the equation has been clarified.  Consensus on a theory of quantum gravity lacks, but tentative theories, such as strings and loops, exist today.  Well defined versions of the equation have been found, and they  are utilized to produce predictions, with a hope of testing them against observation. The equation is  recognized as a basic tool for thinking about the quantum properties of spacetime by a large community of physicists in areas ranging from early cosmology to black hole physics.  The equation has proven an inexhaustible source of inspiration on the path towards understanding the quantum properties of space and time. 

The main reason for this is that the WdW equation has been the icebreaker towards the construction of a quantum theory which does not presuppose a single spacetime.   The physics community underwent a major reshaping of its way of thinking when it begun to utilize general relativity: from a conception of physics as the theory of what happens \emph{in space and time}, to a new conception, where the theory describes also of what happens  \emph{to space and time}. This step has been difficult in the classical framework, where questions like whether gravitational waves were physical or gauge, or concerning the nature of the Schwarzschild singularity, have confused the community for decades.  But it has taken longer, and in fact is still confusing the community, for the quantum regime.   Quantum mechanics and general relativity, taken together, imply the possibility of quantum superposition of different spacetimes.
The WdW equation, which is based on a wave function $\Psi[q]$ over geometries, has offered the first conceptual scheme for dealing with this physical possibility  For this reason it is a milestone.   

Today's tentative quantum gravity theories take  for granted the absence of a single predetermined smooth spacetime at all scales.  The breakthrough opening the path to this thinking happened at the Raleigh-Durham airport 1965 meeting.

\section{From Einstein-Hamilton-Jacobi to Einstein-Schr\"odinger}

In the following I make no attempt at a full review of the vast literature where the WdW equation has been discussed, utilized, and has left its mark. Instead, I  focus on the main issues the equation has raised, on what we have learned from it, and on the possibility it has opened for describing quantum spacetime. 

I start by illustrating the path leading to the WdW equation via Peres' Hamilton-Jacobi formulation of general relativity \cite{Peres1962}, in a bit more detail than above. This is relevant not only for history, but especially  because it clarifies the meaning of the WdW equation, shedding light on the confusion it raised.  Some difficulties of the WdW equation are more apparent than real. They have nothing to do with quantum mechanics; they stem from the application of the Hamilton-Jacobi language to a generally covariant context.

Einstein wrote general relativity in terms of the Lorentzian metric $g_{\mu\nu}(\vec x, t)$, where $\mu,\nu=0,...,3$ are spacetime indices and $(\vec x, t)$ are coordinates on spacetime.  Peres' starting point was the Arnowit-Deser-Misner change of variables \cite{Arnowitt:1962}
\begin{eqnarray}
q_{ab}=g_{ab},\ \ 
N_a = g_{ao},\ \ 
N=\sqrt{-g^{oo}}.
\end{eqnarray}
where $a,b=1,2,3$. $N$ and $N^a=q^{ab}N_b$ are the ``Lapse"\index{Lapse function} and ``Shift"\index{Shift function} functions, and $q^{ab}$ is the inverse of $q_{ab}$.  The interest of these variables is that the Lagrangian does not depend on the time derivatives of Lapse\index{Lapse function} and Shift. Therefore the only true dynamical variable is the three metric $q_{ab}(\vec x)$, the Riemannian metric of a $t=constant$ surface. The canonical hamiltonian vanishes, as is always the case for systems invariant under re-parametrization of the lagrangian evolution parameter, and the dynamics is coded by the two ADM constraints 
\be
   \left(q_{ab}q_{cd}-\frac12 q_{ac}q_{bd}\right)p^{ac}p^{bd}-\det{q}\, R[q]=0  \label{constraint1}
\ee
and
\be
  D_a \, p^{ab}=0  \label{constraint2}, 
\ee
on the three metric $q$ and its conjugate momentum $p$. ($D_a$ is the covariant derivative of the 3-metric.)

The dynamics can be written in Hamilton-Jacobi form by introducing the  Hamilton-Jacobi  function $S[q]$, which is a functional of $q_{ab}(\vec x)$ and demanding that this satisfies the two equations above with $p^{ab}$ replaced by the functional derivative $\delta S[q]/\delta q_{ab}$. The first equation gives  \eqref{peres}, while the second can be rather easily shown \cite{Higgs:1958mh} to be equivalent to the request  
\be
S[q]=S[\tilde q]
\label{six}
\ee
for any two 3d metrics $q$ and $\tilde q$ related a 3d change of coordinates. Solving \eqref{peres} and \eqref{six} amounts to solve the Einstein equations.  To see how, say we have found a family of solutions $S[q,q^0]$ parametrized by a 3-metric $q^0_{ab}(\vec x)$. Then we can define the momenta 
\be
p_0^{ab}[q,q^0](\vec x)=\frac{\delta S[q,q^0]}{\delta q^0_{ab}(\vec x)}  \label{solution}
\ee
which are thus nonlocal functions of $q$ and $q^0$. For any choice of $q^0$ and $p_0$ satisfying \eqref{constraint1} and \eqref{constraint2}, there exists a spacetime which is a solution of Einstein's equations and for any such solution, any field $q$ satisfying  
\be
p_0^{ab}[q,q^0](\vec x) =p_0^{ab}(\vec x).  
\ee
is the metric of a spacelike 3d surface imbedded in this spacetime.  Since the $q^0$ and $p_0$ are related by equations \eqref{constraint1} and \eqref{constraint2}, the last equation does not determine $q$ uniquely: the different solutions correspond to different spacelike surfaces in spacetime, and different coordinatization of the same; in fact, all of them.   Therefore the solution to the Einstein-Hamilton-Jacobi system \eqref{peres}-\eqref{six} provides in principle the full solution of the Einstein's equations.  These results follow from a simple generalization of standard Hamilton-Jacobi theory. 

This was the starting point of Bryce DeWitt.  Now recall that in his milestone 1926 article \cite{Schrodinger:1926fk}, Schr\"odinger introduced (what is called today) the Schr\"odinger equation for the hydrogen atom by taking the Hamilton-Jacobi equation of an electron in a Coulomb potential:
\be
\frac{\partial S(\vec x, t)}{\partial t}+\frac1{2m}\frac{\partial S(\vec x, t)}{\partial \vec x}\cdot \frac{\partial S(\vec x, t)}{\partial \vec x}+\frac{e^2}{|\vec x|}=0
\ee
and replacing derivatives with ($-i\hbar$ times) derivative operators:
\be
\left[-i\hbar\frac{\partial }{\partial t}-\frac{\hbar^2}{2m}\frac{\partial }{\partial \vec x}\cdot \frac{\partial }{\partial \vec x}+\frac{e^2}{|\vec x|}\right]\psi(\vec x, t)=0. 	\label{atom}
\ee
In a shortly subsequent article \cite{Schrodinger1926}, Schr\"odinger offers a rationale for this procedure: the eikonal approximation of a wave equation, which defines the geometrical optic approximation where wave packets follow definite trajectories, can be obtained by the opposite procedure.  If we interpret classical mechanics as the eikonal approximation to a wave mechanics, we can guess the wave equation by this procedure.  Given the immense success of Schr\"odinger's leap, trying the same strategy for gravity is obviously tempting.  This is what led DeWitt and Wheeler to equation \eqref{WdW}. Next to it, the second Hamilton-Jacobi equation, following from \eqref{constraint2}, remains unaltered:
\be
  D_a \, \frac{\delta}{\delta q_{ab}}\Psi[q]=0  \label{qconstraint2}, 
\ee
and, as before, can be simply shown to be equivalent to the requirement that $\Psi[q]=\Psi[q']$ if $q$ and $q'$ are related by a 3d coordinate transformation.  That is, the wave function is only a function of the ``3-geometry", namely the equivalence class of metrics under a diffeomorphism, and not of the specific coordinate dependent form of the $q_{ab}(\vec x)$ tensor.

The Schr\"odinger equation \eqref{atom} gives, in a sense, the full dynamics of the electron in the hydrogen atom. Similarly, one expects the WdW equation \eqref{WdW}, properly understood and properly defined, to give the full dynamics of quantum gravity.  

\section{Physics without background time}

The immediately puzzling aspect of the WdW equation, and the one that has raised the largest confusion, is the absence of a time variable in the equation.  This has been often wrongly attributed to some mysterious quantum disappearance of time.  But things are  simpler: the disappearance of the time variable is already a feature of the \emph{classical} Hamilton-Jacobi formulation of general relativity. It has nothing to do with quantum mechanics. It it is only a consequence of the peculiar manner in which evolution is described in general relativity. 

In Newtonian and special relativistic physics the time variable represent the reading of a clock. It is therefore a quantity to which we associate a well determined procedure of measurement.  Not so in general relativity, where the reading of a clock is not given by the time variable $t$, but is instead expressed by a line integral depending on the gravitational field, computed along the clock's worldline $\gamma$, 
\be
T=\int_\gamma\sqrt{g_{\mu\nu}dx^\mu dx^\nu}.
\label{T}
\ee 
The coordinate $t$ in the argument of $g_{\mu\nu}(\vec x, t)$ which is the evolution parameter of the Lagrangian and Hamiltonian formalisms has no direct physical meaning, and can be changed freely.  Such change in the manner evolution is described is is not a minor step. Einstein's wrote that the biggest difficulty he had to overcome in order to find general relativity was to understand ``the meaning of the coordinates".\footnote{``Why were a further seven years required for setting up
the general theory of relativity? The principal reason is that one does not free oneself so
easily from the conception that an immediate physical significance must be attributed to
the coordinates." Albert Einstein, in \cite{Einstein1949}.}  The physical prediction of general relativity, which can be directly tested against experience are not given by the evolution of physical quantities in the coordinate $t$, but, rather, in the relative evolution of physical quantities with respect to one another. This is why using the time variable $t$ is not required for making sense of general relativity.  

To clarify this crucial point, consider two clocks on the surface of the Earth; imagine one of them is thrown upward, and then falls back down near the first. The reading of the two clocks, say $T_1$ and $T_2$, initially the same, will then differ.  Given the appropriate initial data, general relativity allows us to compute the value of $T_1$ when the second clock reads $T_2$, or viceversa. Does this describe the evolution of\,  $T_1$ in the ``time" $T_2$, or, rather, the evolution of $T_2$ in the ``time" $T_1$? 

The question is clearly pointless: general relativity describes the \emph{relative} evolution of the two variables $T_1$ and $T_2$, both given by \eqref{T} but computed along different worldliness. The two ``times" $T_1$ and $T_2$ are on the same footing.   The example shows that general relativity describes the \emph{relative} evolution of variable quantities with respect to one another, and not the \emph{absolute} evolution of variables in time. 

Mathematically this is realized by parametrizing the motions. Instead of using $T_1(T_2)$ or $T_2(T_1)$ to describe evolution, the theory uses the parametric form $T_1(t),T_2(t)$, where $t$ is an arbitrary parameter, which can be chosen freely.  The coordinates $(t,\vec x)$ in the argument of the gravitational field $g_{\mu\nu}(\vec x, t)$ are arbitrary parameters of this sort. 

This manner of describing evolution is more general than giving the evolution in a preferred time parameter.  The formal structure of dynamics can be generalized to this wider context. This was early recognized by Dirac \cite{Dirac2007} and the corresponding generalized formulation of mechanics has been discussed by many authors in several variants (see for instance \cite{Hanson1976,Rovelli:2004fk}), often under the  deceptively restrictive denomination of ``constrained system dynamics".   ``Constrained system dynamics" is not the dynamics of special systems that have constraints: it is a generalization of dynamics which avoids the need of picking one of the variables and treating it as the  special, independent, evolution parameter. In the corresponding generalization of Hamilton-Jacobi theory, parameter time does not appear at all, because the Hamilton-Jacobi theory gives the relation between observable variables directly. 

Thus, the reason that the coordinate time variable $t$ does not show up in the WdW equation is not at all mysterious after all.  It is the same reason for which coordinates do not show up in the physical predictions of classical general relativity: they have no physical meaning, and the theory can well do without.  

In particular, the absence of $t$ in the WdW equation does not imply at all that the theory describes a frozen world, as unfortunately often suggested. One can pick a function of the gravitational field, or, more realistically, couple a simple system to the gravitational field, and use it as a physical clock, to coordinatize evolution in a physically relevant manner.   A common strategy in quantum cosmology, for instance, is to include a scalar field $\phi(\vec x,t)$ in the system studied, take the approximation where $\phi(\vec x,t)$ is constant in space $\phi(\vec x,t)=\phi(t)$ and give it a simple dynamics, such as a linear growth in proper time.  Then the value of $\phi$ can be taken as a ``clock" --it coordinatizes trajectories of the system-- and the WdW wave function $\Psi[q,\phi]$ can be interpreted as describing the evolution of 
$\Psi[q]$ in the physical variable $\phi$.

The full structure of quantum mechanics in the absence of a preferred time variable has been studied by several authors. Reference \cite{Rovelli:2004fk} is my own favorite version. It can be synthesized as follows. The variables that can interact with an external system (``partial observables") are represented by self-adjoint operators $A_n$ in an auxiliary Hilbert space $\cal K$ where the WdW operator $C$ is defined.  The WdW equation
\be
    C\Psi=0
    \label{C}
\ee
defines a linear subspace $\cal H$, and we call $P:{\cal H}\to{\cal H}$ the orthogonal projector. If zero lies in the continuous spectrum of $C$, then $\cal H$ is not a proper subspace of $\cal K$. It is a generalized subspace namely a linear subspace of a suitable completion of $\cal K$ in a weak norm. In this case, $\cal H$ \emph{still} inherits a scalar product from $\cal K$: this can be defined using various techniques, such as spectral decomposition of $\cal K$, or group averaging.  $P:{\cal H}\to{\cal H}$ is still defined, it is not a projector, as $P^2$ diverges, but we still have $CP=PC=0$.   If $A_n$ is a complete set of commuting operators (in the sense of Dirac) and $|a_n\rangle$ a basis diagonalizing them, then 
\be
W(a';a)=\langle a'_n|P|a_n\rangle
\label{Paa}
\ee
is the amplitude for measuring $\{a'_n\}$ after $\{a_n\}$ has been measured, from which transition probabilities can be defined by properly normalizing. 

This formalism is a generalization of standard quantum mechanics. It reduces to the usual case if 
\be
    C=i\hbar\frac{\partial}{\partial t}+H
\ee
where $H$ is a standard Hamiltonian.  In this case, if $\{q,t\}$ is a complete set of quantum numbers, then 
\be
W(q',t';q,t)=\langle t',q'|P|a,t\rangle
\ee
turns out to be the standard propagator
\be
W(q',t';q,t)=\langle q'|e^{-\frac{i}{\hbar} H t}|q\rangle
\ee
which has the entire information about the quantum dynamics of the system. Thus, in general the WdW equation is just a generalization of the Schr\"odinger equation, to the case where a preferred time variable is not singled out. Equation \eqref{WdW} is the concrete form that \eqref{C} takes when the dynamical system is the gravitational field.  

The first merit of the equation written by Wheeler and DeWitt in 1965 is therefore that it has opened the way to the generalization of quantum theory needed for understanding the quantum properties of our general covariant world. 

The world in which we live is, as far as we understand, well described by a general covariant theory.  In principle, any quantity that we can observe and measure around us can be represented by an operator $A_n$ on $\cal K$, and all dynamical relations predicted by physics can be expressed in terms of the transition amplitudes \eqref{Paa}. In practice, setting up a concrete realization of this framework is complicated.  To see where we are today along the path of making sense of the quantum theory formally defined by the WdW equation, and to describe the legacy of the equation, I now turn to some specific current approach to quantum gravity. 

\section{WdW in Strings}

String theory developed starting from conventional quantum field theory and got in touch with general relativity only later. Even in dealing with gravity, string theory was for sometime confined to perturbations around Minkowski space, where the characteristic features of general relativity are not prominent. For this reason, the specific issues raised by the fact that spacetime is dynamical have not played a major role in the first phases of development.  But major issues cannot remain long hidden, and  at some point string theory has begun to face dynamical aspects of spacetime. In recent years, the realization that in the world there is no preferred spacetime has impacted string theory substantial and the string community has even gone to the opposite extreme, largely embracing more or less precise holographic ideas, where bulk spacetime is not anymore a primary ingredient.  In this context the WdW equation has reappeared in various ways in the context of the theory.  

For instance, in AdS/CFT setting, a constant radial coordinate surface can be seen as playing the role of the ADM constant time surface, the quantisation of the corresponding constraint of the bulk gravity theory gives a WdW equation and its  Hamilton-Jacobi limit turns out to admit an interpretation as a renormalization group equation for the boundary CFT \cite{Lifschytz2000,Papadimitriou2004}. 

In fact, the full AdS/CFT correspondence can be seen as a realization of a WdW framework:  the correspond critically relies on the ADM Hamiltonian being a pure boundary term, since its ``bulk" part is  pure gauge. See \cite{Marolf2009} and, for a direct attempt to derive the AdS/CFT correspondence from the WdW equation, \cite{Freidel2008}.  See also the discussion on dS/CFT, where the ``wave function of the universe" seen as a functional of 3-metrics, just as in canonical general relativity plays a major role \cite{Maldacena2003}.

There is a more direct analog to the WdW equation in the foundation of string theory.  In its ``first quantisation", string theory can be viewed as a two-dimensional field theory on the world-sheet. The action can be taken to be the Polyakov action, which is generally covariant on the world-sheet, and therefore the dynamics is entirely determined by the constraints as in the case of general relativity.  In fact, the two general relativity constraints  \eqref{constraint1} and \eqref{constraint2} have a direct analog in string theory as the Virasoro constraints
\be
  \Pi^2-|\nabla X|^2=0
\ee
and 
\be
  \nabla X\cdot\Pi=0
\ee
where $X$ are the coordinates of the string and $\Pi$ its momentum.  As in general relativity, the second equation implements the invariance under 1-dimensional spatial change of coordinates on the world sheet, while the first is a ``hamiltonian constraint" which codes its dynamics. The left- and right-moving null combinations (associated with $x \pm t$) are  the left- and right-moving Virasoro algebras.   However, the standard quantization of the string is obtained in a manner different from the way the WdW is derived.  The two equations combined in right and left null combinations and Fourier transformed give the Virasoro operators $L_n$, of which only the $n\ge1$ components are imposed as operator equations on the quantum states.  The expectation value of the $L_n$ for negative $n$ vanishes on physical states, so the full constraint is still recovered in the classical limit (in a way similar to Gupta-Bleuler).   In addition, states are required to be eigenstates of $L_0$, with eigenvalue given by the central charge of the corresponding conformal field theory. 

The reason for these choices is that the theory is required to make sense in the target space and respect its Poincar\'e invariance.  But imposing all the constraints $L_n=0$ would be inconsistent, because of the non triviality of the $L_n$'s algebra: a warning about naive treatment of the hamiltonian constraints that must be kept in mind also in the case of general relativity.

\section{WdW in Loops}

The line of research where the WdW equation has had the largest impact is Loop Gravity (LQG), currently the most active attempts to develop a specific theory of quantum gravity addressing  the possibility of quantum superposition of spacetimes. 

LQG, indeed, is born from a partial solution of the WdW equation, and its structure is based on this partial solution. General relativity can be rewritten in terms of variables different from the  metric  used by Einstein.  These, developed by Abhay Ashtekar in the late eighties, are the variables of an $SU(2)$ gauge theory, namely (in the hamiltonian framework) an $SU(2)$ connection $A_a$ and its``electric field" conjugate momentum $E^a$ \cite{Ashtekar:1986yd}.  Rewritten in terms of these variables, the hamiltonian constraint of general relativity reads
\be
    C=F_{ab}E^aE^b =0
\ee 
and the WdW equation takes the simpler form 
\be
    F_{ab}\frac{\delta }{\delta A_a}\frac{\delta }{\delta A_b}\Psi[A]=0. 
    \label{AWdW}
\ee 
Remarkably, we know a large number of solutions of this equation. These were first discovered using a lattice discretization by Ted Jacobson and Lee Smolin \cite{Jacobson:1987qk}, and can be constructed, in the continuum, as follows.  Choose a loop $\gamma=S_1\to R^3$, namely a closed line in space and consider the trace of the holonomy of $A$ along this loop, namely the quantity
\be
  \Psi_\gamma[A]=Tr\,P e^{\oint_\gamma A}
\ee 
where $P$ indicates the standard path ordered exponentiation. It turns out that $\Psi_\gamma[A]$ is a solution of the Ashtekar-WdW equation \eqref{AWdW}, if the loop has no self-intersection (if $\gamma$ is injective) \cite{Rovelli:1987df,Rovelli:1989za}.  A simplified derivation of this result is to observe that the functional derivative of a holonomy vanishes in the space points $\vec x$ outside the loop and is otherwise proportional to the tangent $\dot\gamma^a=d\gamma^a(s)/ds$ to the loop: 
\be
    \frac{\delta \Psi_\gamma}{\delta A_a(\vec x)}=
    \oint_\gamma ds\ \dot\gamma^a(s)\,A_a(\gamma(s))\,\delta^3(\gamma(s),\vec x)\, P e^{\int_\gamma A}
\ee 
where a trace is understood and the path order integral starts at the loop point $\vec x$.  The left hand side of \eqref{AWdW}, is therefore proportional to $F_{ab}\ \dot\gamma^a\,\dot\gamma^b$, which vanishes because of the antisymmetry of $F_{ab}$ if the loop has a no self-intersection. If it has intersections, in the intersection point there are two different tangents and mixed terms do not cancel.  Thus, loop states $\Psi_\gamma$ without intersections are exact solutions of the WdW equation. 

Non-intersecting loop-states alone do not describe a realistic quantum space, because they are eigenstates of the volume $V=\det{q_{ab}}$ with vanishing eigenvalue. (Indeed, $V^2\sim \epsilon_{abc} E^a E^b E^c\sim \epsilon_{abc} \dot\gamma^a \dot\gamma^b \dot\gamma^c=0$).  Therefore intersections play a role in the theory \cite{Brugmann:1991fk}. But acting on a loop state \emph{with} intersections, the Ashtekar-WdW operator \emph{acts non trivially only at the intersection point.}  This is the basis fact underpinning LQC. 

The loop representation of quantum general relativity can formally be obtained by moving from the connection basis $\Psi[A]=\langle A|\Psi \rangle$ to a basis formed by loop states with intersections. An orthonormal basis of such loop-based states with intersection is given by the spin network basis \cite{Rovelli1995b}, which provides today the standard basis on which the theory is defined.  In this basis, the WdW operator acts only at intersections, which are called the ``nodes" of the network.  A rigorous and well-defined definition of the WdW operator in this representation has been given by Thomas Thiemann, and is constructed and studied in detail in his book \cite{ThiemannBook}. 

Simplified versions of this operator are heavily used in the field of Loop Quantum Cosmology, namely the application of loop-gravity results to quantum cosmology \cite{Ashtekar:2013hs}.  Loop quantum cosmology is producing some tentative preliminary predictions about possible early universe quantum gravity effects on the CMB (see for example \cite{Agullo2012a}).  If these were verified, the result would be of major importance and WdW would have been a key ingredient.

\section{WdW and path integrals}

Since its earliest days \cite{Misner:1957fk}, the inspiration for the search for a good quantum theory of quantum gravity has oscillated between the canonical WdW framework and the covariant framework provided by a ``path integral over geometries"
\be
 Z=  \int D[g] \ e^{i\int \sqrt{g}R[g]}.
 \label{integral}
\ee
It is hard to give this integral a mathematical sense, or  to use it for computing transition amplitudes within some approximation scheme, but the formal expression \eqref{integral} has provided an intuitive guidance for constructing the theory.  Formally, the path integral is  related to the WdW equation, in same manner in which the Feynman path integral that defines the propagator a non relativistic particle is a solution of the Schr\"odinger equation.  This relation has taken a particularly intriguing form in the context of the Euclidean quantum gravity program, developed by Hawking and his collaborators  \cite{Hawking:1980gf}, where the wave functional 
\be
      \Psi[q]  \int_{\partial g =q} D[g] \ e^{-\int \sqrt{g}R[g]}
\ee
can is shown by some formal manipulations to be a solution of the WdW equation.  Here the integration is over \emph{euclidean} 4-metrics, inducing the 3-metric $q$ on a 3d boundary. The construction is at the basis of beautiful ideas such as the Hartle-Hawking ``no boundary" definition of a wave function for cosmology \cite{Hartle:1983ai,Halliwell:1984eu}.   

The oscillation between canonical and covariant methods is well known in fundamental theoretical physics, and is not peculiar of quantum gravity.  The two approaches have complementary strengths: the hamiltonian theory captures aspects that are easily overlook ed in the covariant language, especially in the quantum context, while the lagrangian framework allows symmetries to remain manifest,  is  physically far more transparent, and leads to more straightforward calculations techniques. 

In the loop context, the difficulties of dealing with the WdW equation have pushed a good part of the  community to adopt alternative, covariant methods for computing the transition amplitudes \eqref{Paa}, which makes use of the so called ``spinfoam" techniques \cite{Rovelli}, a sum-over-paths technique of computing amplitudes between spin network states which can be seen as a well-defined version of equation  \eqref{integral}.

After all, the initial unhappiness of Bryce DeWitt with an equation entangled within the complexities of the Hamiltonian formalism and having the bad manners of breaking the manifest covariance between space and time, were not unmotivated.  The WdW equation is not necessarily the best manner for actually defining the quantum dynamics and computing transition amplitudes in quantum gravity. 

But it remains the equation that has opened the world of background independent quantum gravity,  a unique source of inspiration, and a powerful conceptual tool that has forced us to understand how to actually make sense of a quantum theory of space and time.


\begin{thebibliography}{10}

\bibitem{Peres1962}
A.~Peres, ``{On Cauchy's problem in General Relativity},'' {\em Nuovo Cimento},
  vol.~26, p.~53, 1962.

\bibitem{DeWitt:1967yk}
B.~S. DeWitt, ``{Quantum Theory of Gravity. 1. The Canonical Theory},'' {\em
  Phys. Rev.}, vol.~160, pp.~1113--1148, 1967.

\bibitem{Wheeler:1968}
J.~A. Wheeler, ``{Superspace and the nature of quantum geometrodynamics},'' in
  {\em Batelles Rencontres} (B.~S. DeWitt and J.~A. Wheeler, eds.), Bennjamin,
  New York, 1968.

\bibitem{Dewitt-Morette2011}
C.~DeWitt-Morette, {\em {The Pursuit od Quantum Gravity}}.
\newblock Springer-Verlag, 2011.

\bibitem{Arnowitt:1962}
R.~L. Arnowitt, S.~Deser, and C.~W. Misner, ``{The dynamics of general
  relativity},'' {\em Gravitation: Introduction to Current Research}, p.~227,
  1962.

\bibitem{Higgs:1958mh}
P.~W. Higgs, ``{Integration of Secondary Constraints in Quantized General
  Relativity},'' {\em Phys. Rev. Lett.}, vol.~1, pp.~373--374, 1958.

\bibitem{Schrodinger:1926fk}
E.~Schr\"{o}dinger, ``{Quantisierung als Eigenwertproblem (Erste
  Mitteilung)},'' {\em Annalen der Physik}, vol.~79, pp.~361--76, 1926.

\bibitem{Schrodinger1926}
E.~Schr\"{o}dinger, ``{Quantisierung als Eigenwertproblem (Zweite
  Mitteilung)},'' {\em Annalen der Physik}, vol.~79, pp.~489--527, 1926.

\bibitem{Einstein1949}
A.~Einstein, ``{Autobigraphical Notes},'' in {\em Albert Einstein:
  Philosopher-Scientist} (P.~Schilpp, ed.), pp.~2-- 94, London: Open Court: La
  Salle/Cambridge University Press, 1949.

\bibitem{Dirac2007}
P.~A.~M. Dirac, ``{Generalized Hamiltonian Dynamics},'' {\em Proc. R. Soc.
  Lond.}, vol.~246, pp.~326--332, 1958.

\bibitem{Hanson1976}
A.~Hanson, T.~Regge, and C.~Teitelboim, {\em {Constrained Hamiltonian
  Systems}}.
\newblock Roma: Accademia Nazionale dei Lincei, 1976.

\bibitem{Rovelli:2004fk}
C.~Rovelli, {\em {Quantum Gravity}}.
\newblock Cambridge, U.K.: Cambridge University Press, 2004.

\bibitem{Lifschytz2000}
G.~Lifschytz and V.~Periwal, ``{Schwinger-Dyson = Wheeler-DeWitt: gauge theory
  observables as bulk operators},'' {\em Journal of High Energy Physics},
  vol.~2000, pp.~026--026, Apr. 2000.

\bibitem{Papadimitriou2004}
I.~Papadimitriou and K.~Skenderis, ``{AdS/CFT correspondence and Geometry},''
  p.~30, Apr. 2004.

\bibitem{Marolf2009}
D.~Marolf, ``{Unitarity and holography in gravitational physics},'' {\em
  Physical Review D}, vol.~79, p.~044010, Feb. 2009.

\bibitem{Freidel2008}
L.~Freidel, ``{Reconstructing AdS/CFT},'' p.~34, Apr. 2008.

\bibitem{Maldacena2003}
J.~Maldacena, ``{Non-gaussian features of primordial fluctuations in single
  field inflationary models},'' {\em Journal of High Energy Physics},
  vol.~2003, pp.~013--013, May 2003.

\bibitem{Ashtekar:1986yd}
A.~Ashtekar, ``{New Variables for Classical and Quantum Gravity},'' {\em Phys.
  Rev. Lett.}, vol.~57, pp.~2244--2247, 1986.

\bibitem{Jacobson:1987qk}
T.~Jacobson and L.~Smolin, ``{Nonperturbative Quantum Geometries},'' {\em Nucl.
  Phys.}, vol.~B299, p.~295, 1988.

\bibitem{Rovelli:1987df}
C.~Rovelli and L.~Smolin, ``{Knot Theory and Quantum Gravity},'' {\em Phys.
  Rev. Lett.}, vol.~61, p.~1155, 1988.

\bibitem{Rovelli:1989za}
C.~Rovelli and L.~Smolin, ``{Loop Space Representation of Quantum General
  Relativity},'' {\em Nucl. Phys.}, vol.~B331, p.~80, 1990.

\bibitem{Brugmann:1991fk}
B.~Br\"{u}gmann and J.~Pullin, ``{Intersecting N loop solutions of the
  Hamiltonian constraint of quantum gravity},'' {\em Nucl. Phys. B}, vol.~363,
  pp.~221--244, 1991.

\bibitem{Rovelli1995b}
C.~Rovelli and L.~Smolin, ``{Spin Networks and Quantum Gravity},'' {\em
  arXiv.org}, vol.~gr-qc, 1995.

\bibitem{ThiemannBook}
T.~Thiemann, {\em {Modern Canonical Quantum General Relativity}}.
\newblock Cambridge, U.K.: Cambridge University Press, 2007.

\bibitem{Ashtekar:2013hs}
A.~Ashtekar, ``{Quantum Gravity and Quantum Cosmology},'' {\em Lect. Notes
  Phys.}, vol.~893, pp.~31--56, 2013.

\bibitem{Agullo2012a}
I.~Agullo, A.~Ashtekar, and W.~Nelson, ``{A Quantum Gravity Extension of the
  Inflationary Scenario},'' {\em Phys. Rev. Lett.}, vol.~109, p.~251301, 2012.

\bibitem{Misner:1957fk}
C.~W. Misner, ``{Feynman quantization of general relativity},'' {\em Rev Mod
  Phys}, vol.~29, p.~497, 1957.

\bibitem{Hawking:1980gf}
S.~W. Hawking, ``{The path integral approach to quantum gravity},'' {\em
  General Relativity}, pp.~746--789, 1980.

\bibitem{Hartle:1983ai}
J.~B. Hartle and S.~W. Hawking, ``{Wave Function of the Universe},'' {\em Phys.
  Rev.}, vol.~D28, pp.~2960--2975, 1983.

\bibitem{Halliwell:1984eu}
J.~J. Halliwell and S.~W. Hawking, ``{The Origin of Structure in the
  Universe},'' {\em Phys. Rev.}, vol.~D31, p.~1777, 1985.

\bibitem{Rovelli}
C.~Rovelli and F.~Vidotto, {\em {Introduction to covariant loop quantum
  gravity}}.
\newblock Cambridge University Press, to appear, 2014.

\end{thebibliography}
\end{document}